\documentclass{article}

\usepackage{PRIMEarxiv}

\usepackage[utf8]{inputenc} % allow utf-8 input
\usepackage[T1]{fontenc}    % use 8-bit T1 fonts
\usepackage{hyperref}       % hyperlinks
\usepackage{url}            % simple URL typesetting
\usepackage{booktabs}       % professional-quality tables
\usepackage{amsfonts}       % blackboard math symbols
\usepackage{nicefrac}       % compact symbols for 1/2, etc.
\usepackage{microtype}      % microtypography
\usepackage{lipsum}
\usepackage{fancyhdr}       % header
\usepackage{graphicx}       % graphics

\title{A Combined Finite Element and Finite Volume \\ Method for Liquid Simulation}

\author{
  Tatsuya Koike \\
  Waseda University \\
  \texttt{wassan2356@moegi.waseda.jp} \\
   \And
  Shigeo Morishima \\
  Waseda University \\
  \texttt{shigeo@waseda.jp} \\
  \AND
  Ryoichi Ando \\
  Unaffiliated \\
  \texttt{ryich.ando@gmail.com} \\
}

\usepackage{amsmath,amssymb}
\usepackage{subfigure}
\usepackage{color}
\usepackage{bm}
\usepackage{algorithm}
\usepackage{algpseudocode}
\newcommand{\ry}[1]{\textcolor{black}{#1}}
\newcommand{\fix}[1]{\textcolor{black}{#1}}

\begin{document}
\maketitle

% abstract
%
\begin{abstract}
\small
We introduce a new Eulerian simulation framework for liquid animation that leverages both finite element and finite volume methods. In contrast to previous methods where the whole simulation domain is discretized either using the finite volume method or finite element method, our method spatially \ry{merges} them \fix{together} \ry{using} two types of discretization being tightly coupled on its seams \fix{while enforcing second order accurate boundary conditions at free surfaces.} We achieve our formulation via a variational form \fix{using new} shape functions \fix{specifically designed for this purpose}. By enabling a mixture of the two methods, we \fix{can take advantage of} the best of two worlds. \fix{For example}, finite volume method (FVM) result in sparse linear systems; however, \fix{complexity is encountered when unstructured grids such as tetrahedral or Voronoi elements are used}. Finite element method (FEM), on the other hand, result in comparably denser linear systems, but the complexity remains the same \ry{even if} unstructured elements \ry{are chosen}; \fix{thereby} facilitating spatial adaptivity. In this paper, we propose to use FVM for the majority parts to retain the sparsity of linear systems and FEM for parts where the grid elements are allowed to be freely deformed. An example of this application is locally moving grids. \fix{We show that by} adapting the local grid movement to \fix{an underlying nearly rigid motion, numerical diffusion is noticeably reduced; leading to better preservation} of structural details such as sharp edges, \fix{thin sheets} and spindles \fix{of liquids}.
\end{abstract}
%

%-------------------------------------------------------------------------
\section{Introduction} \label{sec:introduction}
Spatial adaptivity has long been a promising tool for simulating large-scale physics phenomena, including water, smoke, and solids. In general, the key ingredients toward this goal live in the design of discretization of partial differential equations (PDE) on unstructured grids. In the past \fix{decades}, a number of discretization techniques has been proposed such as finite volume method, finite element method, boundary element method and spectrum method.
\par
They have both advantages and disadvantages. For example, spectrum method enable\ry{s} accurate temporal integration since the analytical solutions are known, but are often \fix{limited} by the shape of simulation boundaries such as rectangles or spheres. Boundary element method reduce\ry{s} the degrees of freedom from volumetric $O(n^3)$ to surface area $O(n^2)$, but is only practical if a fast summation of spatially scattered Green's functions is performed. Even though the summation could be accelerated via the fast multipole method (\ry{FMM}), the linear system can not be pre-assembled, indicating that many efficacy-proven pre-conditioners are not \fix{accessible}.
\par
In fluids, finite volume method (FVM) have been historically the most favorable choice because FVM naturally encodes the concept of incompressibility as net zero of momentum flux \fix{over a control volume}. Also, the resulting linear system becomes sparse (e.g., six entries per row for Laplacian on uniform grids). On the other hand, the complexity of FVM rapidly grows when the structure of grids deviates from the uniform cells. In particular, one is required to compute \fix{explicit} Voronoi cells (or power particles) and its associated faces. In contrast, finite element method (FEM) gracefully handle such cases. This is possible because FEM encodes PDEs as volumetric integration over local support domains (a.k.a weak form); in doing this, exact \fix{volumetric} integration of irregular shaped elements is \fix{computed} by a quadrature rule. A quadrature rule requires only the knowledge of mapping from the spatial coordinates to the \fix{unit} coordinate, which is straightforward for tetrahedral or hexahedral grids. At the cost of its versatility, FEM is often more expensive than FVM because \fix{nodes (essentially, the degrees of freedom)} interact with diagonal grid elements. For example, a linearized Laplacian operator in FEM results in $27$ non-zeros per row on uniform grids.
\par
In this \fix{paper}, we \fix{aim to spatially select either} FVM or FEM \fix{where desired}. \fix{A major technical challenge in this end} is \fix{how to design} a \fix{strongly coupled discretization of two methods while minimizing artifacts}. \fix{As an example of our application}, we \fix{propose to populate locally moving grids to reduce} numerical diffusion. \fix{Altogether, our} contributions are summarized as follows.
\par
\begin{itemize}
    \item \fix{Strongly coupled} FEM and FVM \fix{pressure solve}: we propose a new method for calculating pressure \fix{using} finite element method and finite volume method \fix{with second order accurate boundary conditions at free surfaces best satisfied}.
    \item Spatial interpolation \fix{using} MLS: we \fix{apply} a MLS \fix{approximation to interpolate both velocity and level set where} different discretization methods \fix{are spatially adjacent each other}.
    \item \fix{Moving grids}: \fix{we propose to move local grids along the velocity field}, which \fix{we show lessens} numerical diffusion \fix{arising from advection}.
\end{itemize}
We verify the numerical accuracy of our combined finite element and finite volume method \fix{schemes} through various tests. We also compare our simulation results \fix{with/without} moving grids.

\section{Previous Work}
\fix{The history of research of} computational fluid dynamics is long. \fix{In graphics, fluid animation stems} back to Stam \cite{Stam:1999:SF:311535.311548}, \fix{which first incorporated} the semi-Lagrangian advection scheme \fix{in graphics}.
Later, Fedkiw et al. \cite{VSS:2001} proposed a smoke simulation \fix{framework} that \fix{introduces} vorticity confinement. \fix{To this date, one} standard grid-based liquid simulation method is based on combined Marker-And-Cell (MAC) grids and the level set method \cite{HW1965,FM1997}.
Many researchers have extended these methods, such as imposing second order accurate Dirichlet boundary conditions~\cite{FF01,Enright03usingthe} by \fix{taking into account} the distance \fix{from cell centers} to the liquid surfaces.
A broad overview of the history of these fluid simulations can be found in Bridson \cite{bridson:9207783}.

Octree \fix{grids are one of the most prevalent} spatially adaptive methods. In computational mechanics, \fix{some work of this kind adopted} an asymmetric \fix{discretization of Poisson's equation for solving pressure}~\cite{popinet}. Later, smoke and liquid simulations~\cite{ShiYu:2004,Losasso:2004}, \fix{which can dynamically change the structure of} octree were developed and become \fix{widespread} because they can greatly reduce computational run time.
The above methods \fix{were further} improved to be able to handle pressure with second order accuracy on T-junctions~\cite{LOSASSO2006995}.
Such methods can also be \fix{extended to enforce} second order accuracy on solid \cite{fluidcoupling07,2009JCoPh.228.8807N} and liquid surfaces~\cite{Enright03usingthe} \fix{by means of} the cut-cell or ghost fluid methods. However, since these methods are FVM-based, \fix{accurate discretization is not trivial near} T-junction \fix{cells}.
Some researchers proposed a feasible fix to this issue, such as using power diagram \cite{aanjaneya2017} or modifying Poisson discretization \cite{ando2020}.
% Ando and Batty \cite{ando2020} proposed a \fix{feasible fix} to \fix{this issue}, but it is still difficult to deal with grid resolution jump.

Tetrahedral and hexahedral \fix{meshes} for unstructured grids are one of the choices to produce adaptivity. Klingner et al. \cite{Klingner:2006} were the first to use a spatially adaptive tetrahedral mesh for smoke simulation, and Chentanez et al. \cite{Chentanez:2007} extended to liquid simulation.
While these methods suffer from the T-junction problem, Batty et al. \cite{tetrahedra2010} proposed a tetrahedral method that does not \fix{suffer from} this problem, and Voronoi decomposition \cite{Brochu:2010} is another example of a method that can handle boundaries.
However, these discretizations \fix{are inherently complicated}, especially \fix{the Poisson'equation for} pressure solve, and there is still concern \fix{on} the computational cost and memory consumption.

The use of moving grids has been \fix{proposed} by English \cite{english2013} and Fan \cite{EoL2013}. The former is a method to prepare different grids for each moving grid and \fix{stitch} them \fix{together} by using Voronoi cells when constructing a linear system of pressure solve. The latter work was done for solid simulation \fix{but it was not shown for} fluids.
\fix{Aside} from moving grids, hybrid grid-particle methods such as FLuid Implicit Particle (FLIP) \cite{FLIP:2005} and Extended Narrow Band FLIP (EXNBFLIP) \cite{EXNBFLIP:2018} are \fix{feasible alternatives} to avoid numerical diffusion. \fix{They} can be expected to improve the representation of details such as ballistic liquid motion. Combining them with moving grids is expected to provide stable calculations even with large time steps, and \fix{our method may be integrated with such methods.}
\section{Method}
%あと書くならベクトル、行列表記。この場合はFEM圧力の$u^*$を行列表記にして、要素中央で定義しているからconstって明記する。
\fix{Our fluid solver is aimed to numerically solve the incompressible Euler equations}
\begin{equation}
    \frac{\partial \bm{u}}{\partial t} + \bm{u} \cdot \nabla \bm{u} = -\frac{1}{\rho} \nabla p + \bm{g}, \hspace{0.3cm} \nabla \cdot \bm{u} = 0,
    \label{eq:NS}
\end{equation}
where $\bm{u}, t, \rho, p$ and $\bm{g}$ represent velocity, time, density, pressure, and gravity, respectively. In our \fix{discretization}, we set $\rho = 1$ for liquid and $\rho = 0$ for the air. For the boundary conditions, we impose the Dirichlet boundary condition at the liquid-air boundary and the Neumann boundary condition at the liquid-solid boundary \fix{such that} $\bm{n} \cdot \nabla p = 0$ where $\bm{n}$ represents a solid normal. We use an adaptive time step method based on the CFL condition with the CFL number of around 2. \fix{For reasoning of such settings, we refer to Bridson's book \cite{bridson:9207783}.}

We \fix{outline} our algorithm overview in Alg.\ref{alg:overview}. In this section, we first discuss how to discretize the simulation domain by the combined FVM and FEM. Next, we \fix{detail} how to construct a linear system for pressure solve \fix{as well as} how to handle boundary conditions in section \ref{method:section:solver}.
\fix{Finally, we} discuss how to handle advection with moving grids and \fix{how to} interpolate \fix{quantities defined on grids}.

\begin{algorithm}
	\caption{Method Overview}
	\begin{algorithmic}[1]
	    \State Compute time step
	    \State Extrapolate velocity
	    \State Update position of moving grids
	    \State Advect levelset or density by $\bm{u}_{t}$
	    \State Advect velocity by $\bm{u}_{t}$
	    \State Reinitialize levelset
	    \State Pressure solve and project Velocity
	    \State Update velocity of moving grids
	\end{algorithmic} 
	\label{alg:overview}
\end{algorithm}
\subsection{Discretization} \label{method:section:discretize}
\begin{figure}[htb]
  \centering
  \includegraphics[width=\linewidth]{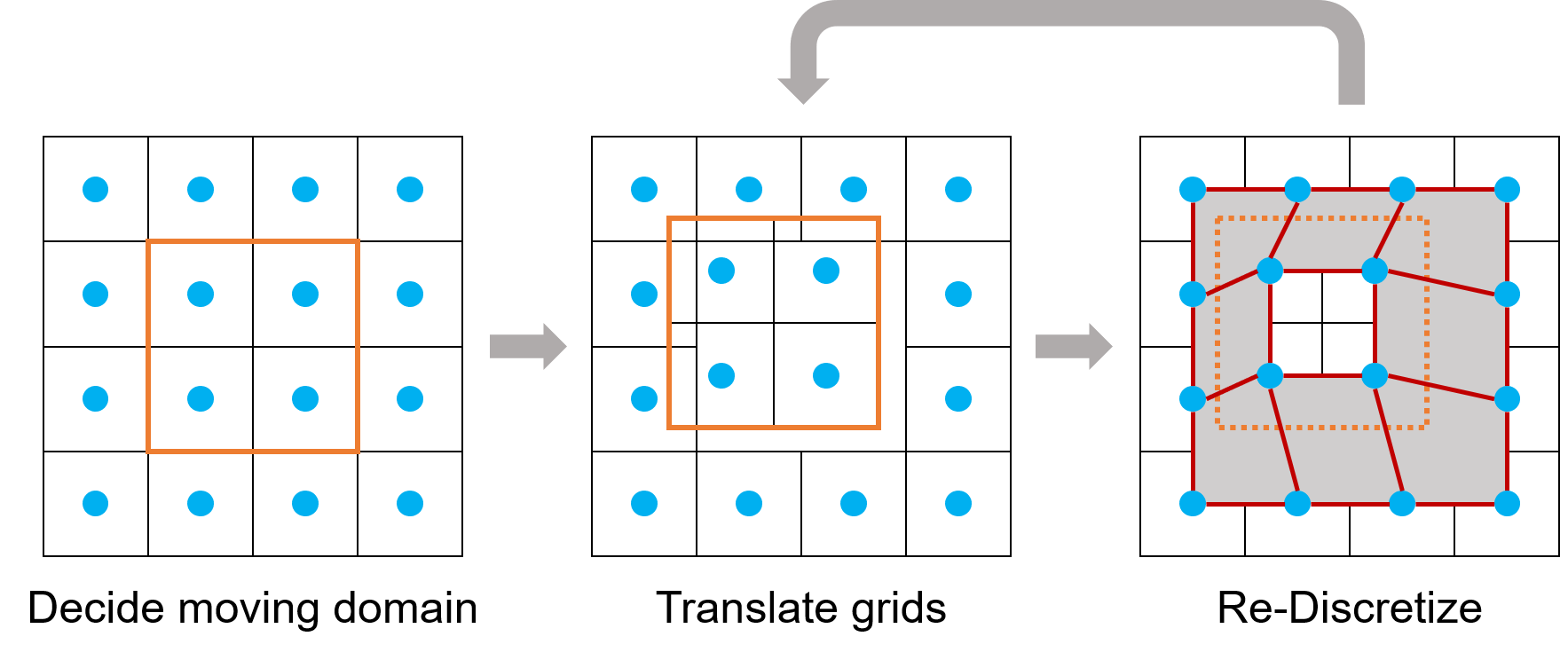}
  \caption[Discretization overview]{\label{fig:discretize}
           Discretization overview of moving grids. We specify the domain of the moving grids in advance (left). During the simulation, the position of the moving grids is updated according to the grid velocity $\bm{u}_g$ (middle). We use FEM (gray area) only at the edges of the moving grids (right).}
\end{figure}
In this paper, we use FVM for the majority part, and FEM is used only \fix{near} the boundary with the moving grids. Figure \ref{fig:discretize} shows an overview of the discretization process. We construct the finite elements using cell-centered points as a node to simplify the pressure solve. When we set the moving grids, we \fix{assign} information on which axis to move. The velocity of the moving grids $\bm{u}_g$ is \fix{set} to be the largest velocity \fix{within the cells}.
\subsection{Pressure solve} \label{method:section:solver}
\begin{figure}[htb]
  \centering
  \includegraphics[width=0.4\linewidth]{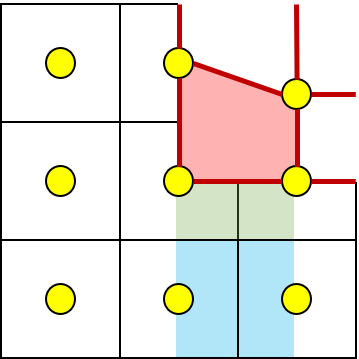}
  \caption[Volume integration and control volume adjustment]{\label{fig:pressure_solve}
           Volume integration and control volume adjustment. The control volume in FVM, colored as blue, is $\Delta x^2$. Red area is intergrated by Gaussian quadrature. The control volume is adjusted near the moving grids like the green area ($0.5 \Delta x^2$).}
\end{figure}
The pressure in fluid simulation can be \fix{cast into} solving the following kinetic energy minimization problem \cite{fluidcoupling07},
\begin{equation}
    \underset{p}{\text{minimize}} \int_\Omega \frac{1}{2} \rho \left\| \bm{u^*} -\frac{\Delta t}{\rho} \nabla p \right\|^2 dV,
\end{equation}
where $\bm{u}^*$ denotes the velocity to be projected, and $\Omega$ represents the entire fluid domain. Since the pressure solve in our method is based on that of FVM and FEM, we will first give a brief explanation of both.

First, we describe the FVM, which accounts for the majority part in our method. To compute pressure, We solve a following linear system \cite{ando2020},
\begin{equation}
    [\nabla]^T [V] [A] [F] [\nabla] \{p\} = [\nabla]^T [V] [A] \{\bm{u}^*\},
    \label{eq:fvm_solver}
\end{equation}
where $[\nabla]^T, [\nabla], [V], [A], $ and $[F]$ respectively denotes a discrete divergence operator, a discrete gradient operator, a control volume of each face, a fluid area fraction and \fix{an} inverse of liquid fraction of each face to achieve second order accuracy for Dirichlet boundary condition~\cite{Enright03usingthe}.

As for the pressure solve in FEM, we \ry{devise the method presented by} Ibayashi et al \cite{Ibayashi2018}. 
We start by \fix{mapping} the spatial coordinate $\bm{x}$ to unit coordinate $\bm{\xi} = [\xi, \eta, \zeta]$ \ry{similarly to} various finite element methods.
In this \ry{setting}, the quantity $q$ at arbitrary coordinates in 3D can be rewritten as follows,
\begin{equation}
    q(\bm{\xi}) = \sum_{i=1}^{8} N_i(\bm{\xi}) q_i,
\end{equation}
where $N_i$ is the trilinear shape function and is defined at the node of each element.
Under this assumption, we construct a linear system as follows,
\begin{equation}
    \sum^n_{e=1} \int_{\Omega_e} [\nabla_e]^T |J| [D] [\nabla_e] \{p_e\} dV_{\xi} = \sum^n_{e=1} \int_{\Omega_e} [\nabla_e]^T |J| \bm{u}^* dV_{\xi},
    \label{eq:fem_solver}
\end{equation}
where $\{p_e\} \in \mathbb{R}^{8 \times 1}$ is the pressure stored on the nodes of $e$th element, and $J$ is the Jacobian matrix. $[\nabla_e]$ is the gradient operator given as,
\begin{equation}
    [\nabla_e]_{ij} = [J^{-1} \nabla_{\xi} N(\bm{\xi})]_i, \hspace{0.5cm} \nabla_{\xi} = \left( \frac{\partial}{\partial \xi},\frac{\partial}{\partial \eta},\frac{\partial}{\partial \zeta} \right)^T.
\end{equation}
$[D]$ is defined as follows,
\begin{equation}
    [D] = \frac{1}{\rho} M_{\phi},
\end{equation}
where $M_\phi$ is the matrix that imposes a second order accuracy on the pressure solve proposed by Ibayashi et al \cite{Ibayashi2018}. We integrate Eq.(\ref{eq:fem_solver}) with an eight-point Gaussian quadrature integration scheme. We use the velocity of each element \fix{interpolated on} Gaussian quadrature points for simplicity.
\par
\fix{If we compare} Eq.(\ref{eq:fvm_solver}) and Eq.(\ref{eq:fem_solver}), we can see that volume factor $[V]$ and second order operator $[F]$ in the FVM correspond to $|J|$ and $[D]$ in the FEM, respectively. Since we have used the same \fix{spatial points} of pressure definition in both methods, the pressure $p$ in each equation is considered to \fix{exactly coincide}. As a result, we can properly compute the pressure by constructing a linear system that combines corresponding matrices. It is important to note that there is an overlap in the integration \fix{domain on} the \fix{seams} between FEM and FVM.
We deal with this problem by adjusting the control volume of the FVM according to the occupancy of the finite elements as shown in the green area of Figure \ref{fig:pressure_solve}. Because the FEM uses Gaussian quadrature interpolation scheme and \fix{mapping it} to \fix{the} unit coordinate, it is difficult to adjust the integration range.

We \fix{project our} second order operator \fix{such that} the linear system \fix{becomes Symmetric} Positive Definite (SPD) matrix that is numerically stable, as \fix{shown by} previous work. The accuracy is therefore \fix{compromised}, but we did not observe any visual artifact by this \fix{projection}. A comparison with the \fix{fully} second order solver is discussed in Section \ref{chap:discussion} on the subject of tilted pools.

Once the pressure is calculated, we have to project the velocity to the divergence-free field, so we subtract the pressure gradient
\begin{equation}
    \{\bm{u}\} = \{\bm{u^*}\} - [F][\nabla]\{p\}.
\end{equation}
For FEM, the velocity is defined at the center of the element, so the subtraction is averaged \fix{from} the calculation points of the Gaussian quadrature interpolation scheme.

\subsection{Interpolation} \label{method:section:interp}
\begin{figure}[htb]
    \centering
    \subfigure{\includegraphics[width=0.4\linewidth]{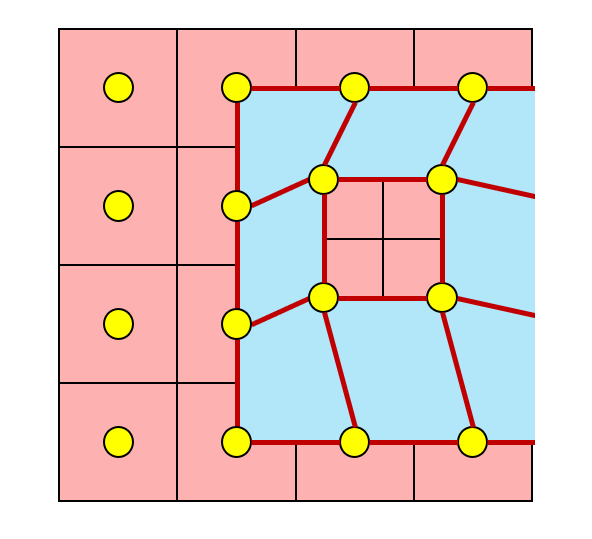}}
    \subfigure{\includegraphics[width=0.4\linewidth]{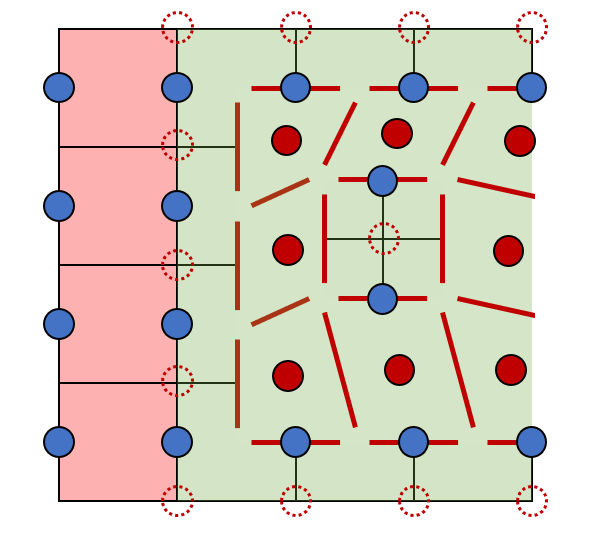}}
    \caption[Interpolation of cell-centered and face-centered values]{\label{fig:interp_centered}
        Our interpolation scheme for cell-centered value (left) and face-centered value (right). Yellow, dark blue, dark red, and the dark hollow red circle represent the location of cell-centered value, face-centered value of FVM, that of FEM, and the nodes where the element does not exist, respectively.}
\end{figure}
\fix{For the regularly distributed points} use trilinear interpolation \fix{both for} the FVM domain and the FEM domain to interpolate the values. \fix{For regions where two methods meet each other}, we use different strategies depending on the location of quantity.

First, \fix{consider} the cell-centered values. Because we share the location of the cell-centered values, we simply switch interpolation methods depending on whether the location where we want to interpolate in the FVM or \fix{the} FEM region. For the FVM domain (\fix{red} colored area of Figure \ref{fig:interp_centered}), we interpolate the values by trilinear interpolation, and for the FEM domain (blue area), we use \fix{element-wise} interpolation with the unit coordinate computed \fix{using Newton's} method.

Unlike the cell-centered values, the location of the face-centered values of FVM and FEM are not shared, \fix{therefore; we sort to} Moving Least Square (MLS) interpolation method for face-centered values near the boundary (green area).
As \fix{suggested} by Ando et al. \cite{ando2020} we perform MLS interpolation based on the following equation,
\begin{align}
    q(\bm{p}) = \begin{bmatrix} p_x & p_y & p_z & 1 \end{bmatrix} (Z^T D Z)^{-1} Z^T D \begin{bmatrix} \{q\}_1 \\ \vdots \\ \{q\}_n \end{bmatrix}, \\
    Z = \begin{bmatrix}
        x_1 & y_1 & z_1 & 1 \\
        \vdots & \vdots & \vdots & \\
        x_n & y_n & z_n & 1 \\
        \end{bmatrix}, \hspace{0.3cm} D = \rm{diag} \begin{pmatrix} \begin{bmatrix} N(\bm{p} - \bm{x}_1, \Delta x) \\ \vdots \\ N(\bm{p} - \bm{x}_n, \Delta x) \end{bmatrix} \end{pmatrix},
\end{align}
where $\bm{p} = [p_x, p_y, p_z], \bm{x}_i = [x_i, y_i, z_i], n,$ and $q$ denotes the query point, the position of the $i$th sample point, the number of sample points, and variable at the discrete point, respectively.
We utilize such an interpolation scheme by using different ways of determining weight $N$ for each FVM and FEM values.
For the FVM weights, we use
\begin{equation}
    N_{FVM} (\bm{r}, h) = \prod^3_{j=1} \rm{max} (1 - \frac{||r_j||}{h}, \epsilon),
\end{equation}
where $\epsilon$ is the safety constant value. This weight is consistent with a simple tri-linear interpolation weight. For FEM, we first compute the unit coordinate $\bm{\xi}$ and use coefficients of the elemental interpolation as the FEM weights. To compute $\bm{\xi}$, we consider virtual elements composed by connecting the defining point of the velocity, that is the center of each element. As shown in the Figure \ref{fig:interp_centered}, in the part where no element is defined, we assume that the element center is at the grid node point, and $\bm{\xi}$ is computed from this virtual elements. If not enough samples are collected, MLS interpolation can be stabilized by collecting samples from the surrounding area with weights $\epsilon$.
\subsection{Advection} \label{method:section:advection}
We use semi-Lagrangian advection scheme \cite{Stam:1999:SF:311535.311548} for both the velocity and the levelset advection. In our method, the defining points of the physical quantities are updated at each simulation step. \fix{For this reason,} we need to pay attention to the location of the physical quantities used in the advection term.
The equation for updating the values by semi-Lagrangian method is as follows
\begin{equation}
    q^*(\bm{x}, t) = q(\bm{x} - \bm{u}(\bm{x} ,t) \Delta t, t)
    \label{eq:semi-lagrangian}
\end{equation}
where $q^*$ denotes the intermediate quantity after the advection, and $q$ denotes the quantity that will be advected. In our method, we can consider that $q$ is the quantity defined on the grid before updated. Fortunately, we can consider Eq.(\ref{eq:semi-lagrangian}) as advection using relative velocity, because it is consistent with the derivation based on the following advection equation,
\begin{equation}
    q^*(\bm{x}, t) = q(\bm{x} -\bm{u}_g \Delta t - (\bm{u}(\bm{x} ,t) - \bm{u}_g) \Delta t, t),
\end{equation}
where the second term of the argument of $q$ represents the moving grid displacement, and the third term represents the advection by relative velocity.
Such a strategy can be considered as the simple version of the advection method proposed in Chimera Grids \cite{english2013}. The difference between our method and theirs is that we do not need to prepare several layers of ghost cells due to the continuity of our interpolation method as described in Section \ref{method:section:interp}.
\section{Results}
We ran some experiments on a Linux machine with AMD Ryzen 9 5950X. We use Eigen library \cite{eigenweb} to solve the linear system, and set the tolerant relative residual to \fix{$10^{-4}$} for pressure solve. We use the PDE-based approach \cite{Russo2000ARO} for the level set reinitialization process. We apply the marching cubes \cite{MarchigCubes1987} for liquid surface extraction, and visualize them by Mitsuba \cite{Mitsuba}. In this section, we will evaluate the accuracy of the combined finite element and finite volume methods, and we show the benefits of simulation using moving grids.
\subsection{Accuracy evaluation}
\begin{figure}[htb]
    \centering
    \subfigure{\includegraphics[width=0.4\linewidth]{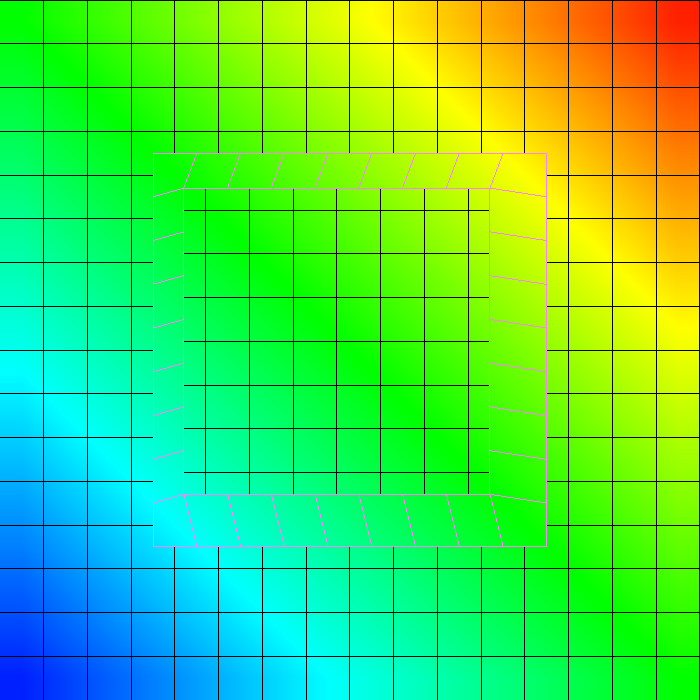}}
    \subfigure{\includegraphics[width=0.4\linewidth]{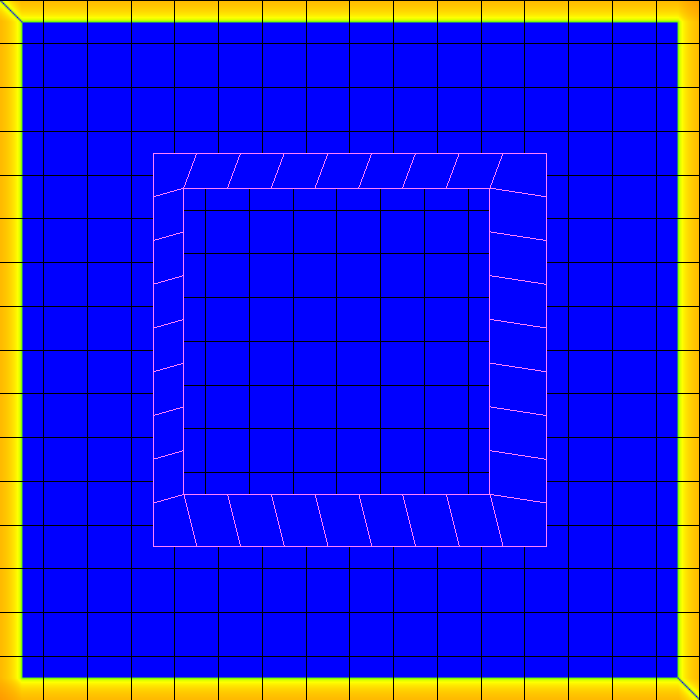}}
    \subfigure{\includegraphics[width=0.4\linewidth]{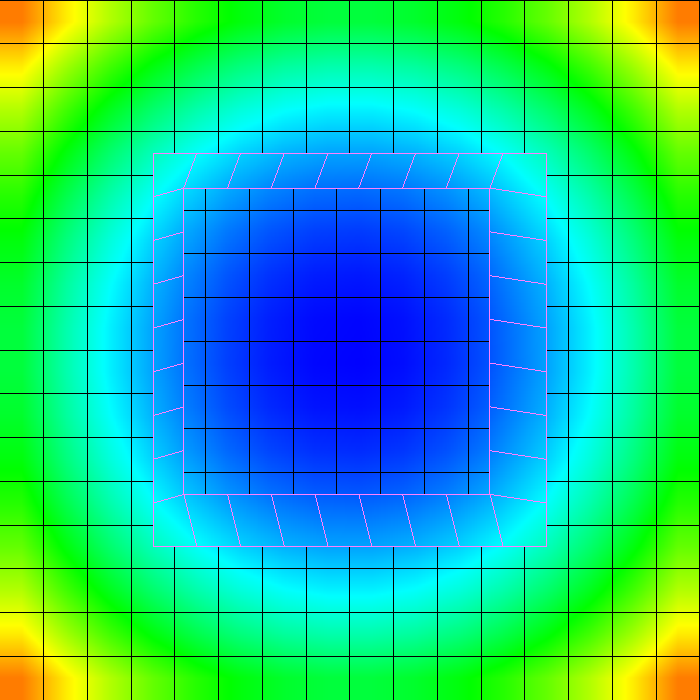}}
    \subfigure{\includegraphics[width=0.4\linewidth]{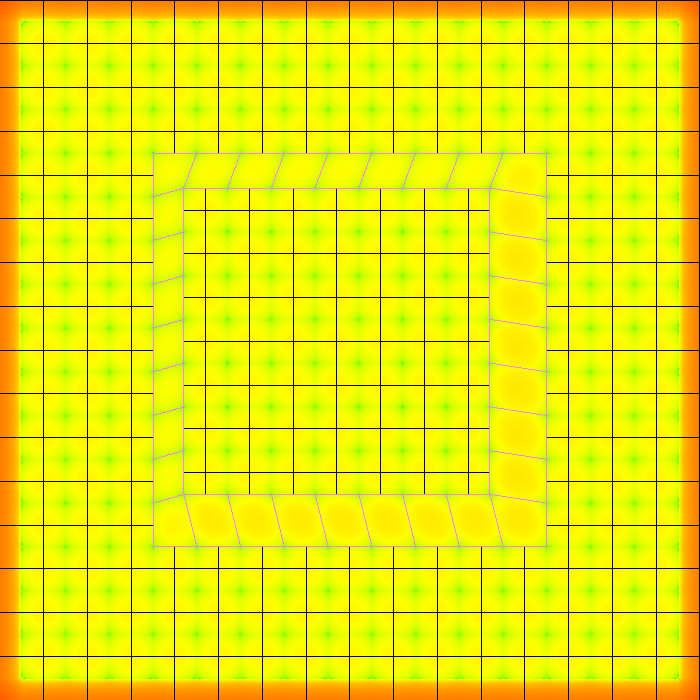}}
    \subfigure{\includegraphics[width=0.85\linewidth]{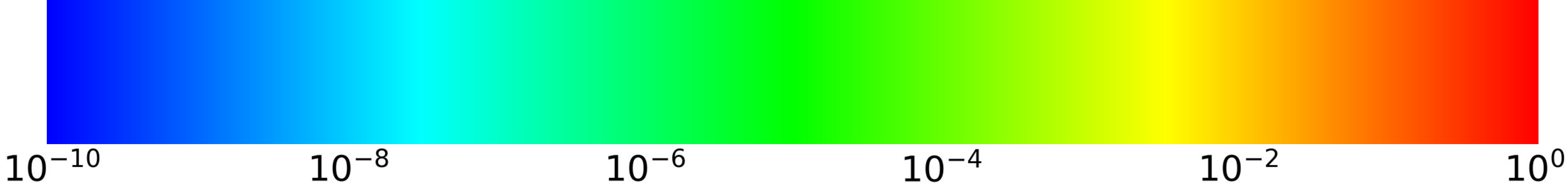}}
    \caption[Results of cell centered value interpolation]{\label{fig:error_cell}
        Interpolation results of cell centered values for linear (top) and quadratic (bottom) fields. Left: reproduced field. Right: error visualization. }
\end{figure}

\begin{figure}[htb]
    \centering
    \subfigure{\includegraphics[width=0.4\linewidth]{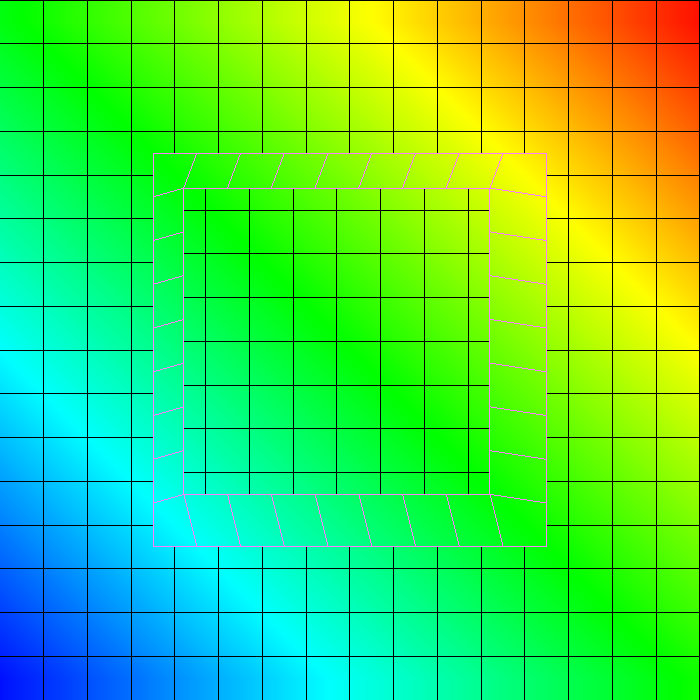}}
    \subfigure{\includegraphics[width=0.4\linewidth]{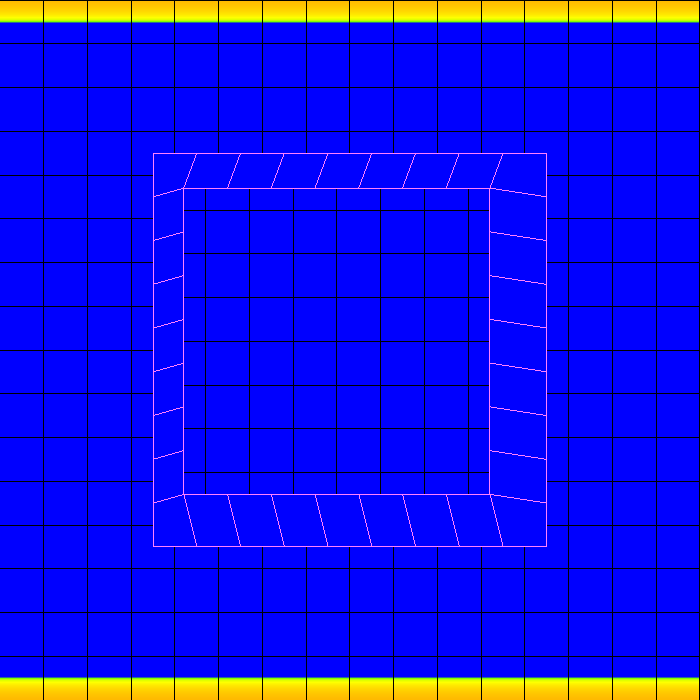}}
    \subfigure{\includegraphics[width=0.4\linewidth]{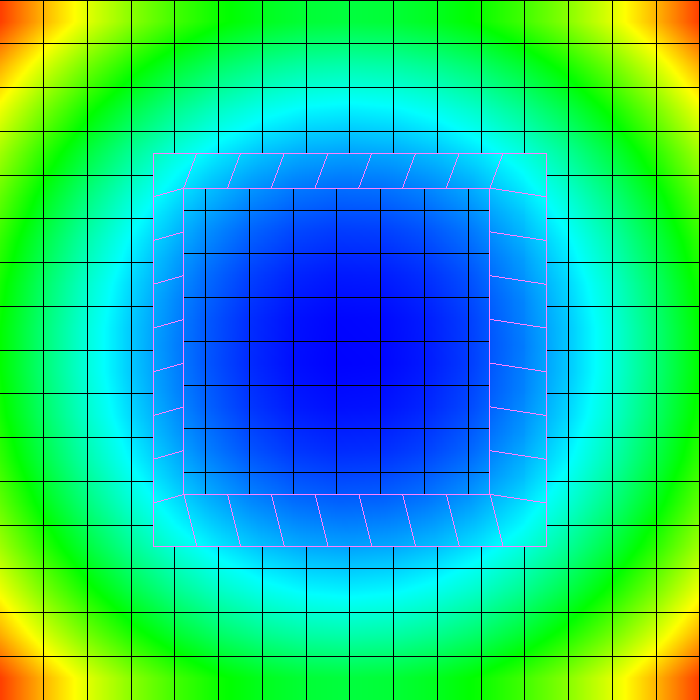}}
    \subfigure{\includegraphics[width=0.4\linewidth]{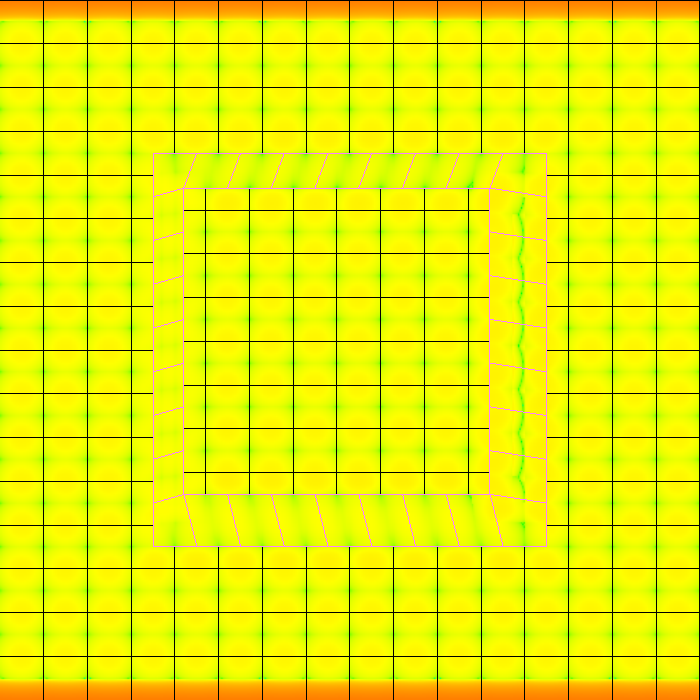}}
    \subfigure{\includegraphics[width=0.85\linewidth]{figures/interp_test/colorbar_with_scale.pdf}}
    \caption[Results of face centered value interpolation]{\label{fig:error_face}
        Interpolation results of face centered values for linear (top) and quadratic (bottom) fields. Left: reproduced field. Right: error visualization. }
\end{figure}

\begin{table}[htb]
  \centering
  \begin{tabular}{|c|l|l|} \hline
    Resolution & Cell centered & Face centered \\ \hline
    $16^2$ & $5.332 \text{e-3} (\text{N / A})$ & $4.349 \text{e-3} (\text{N / A})$ \\
    $32^2$ & $1.333 \text{e-3} (2.00)$ & $1.086 \text{e-3} (2.00)$ \\
    $64^2$ & $3.324 \text{e-4} (2.00)$ & $2.716 \text{e-4} (2.00)$ \\
    $128^2$ & $8.325 \text{e-5} (2.00)$ & $6.726 \text{e-4} (2.01)$ \\
    $256^2$ & $2.047 \text{e-5} (2.02)$ & $1.669 \text{e-5} (2.01)$ \\ \hline
  \end{tabular}
  \caption{$L_{\infty}$ error and accuracy order of our interpolation scheme on quadratic field.}
  \label{tab:interp_error}
\end{table}
We evaluate our interpolation method with linear or quadratic fields, and the results are shown in Figure \ref{fig:error_cell} and Figure \ref{fig:error_face}. We set up a displaced grid and used the FEM around it as well as the discretization of the moving grids. We set the range as a $1 \times 1$ dimensionless field, and each field is represented by the following equation.
\begin{align}
    \text{Linear} & : f(x, y) = 0.5 (x + y) \\
    \text{Quadratic} & : f(x, y) = 2 \left\{(x - 0.5)^2 + (y - 0.5)^2 \right\}.
\end{align}
We convert the errors into the following log scale,
\begin{equation}
    f(x,y)= 1.0 + 0.1 \log_{10} |\text{error}(x,y)|,
\end{equation}
and visualized it in a heat map. Our interpolation method is able to reproduce the linear field \fix{with minimal} errors. Because it is based on trilinear interpolation, there are some errors in the quadratic field, but it decreases \fix{at} the second order as the resolution \fix{doubles}. (Table \ref{tab:interp_error}).
\par
We also tested the accuracy of our pressure solve on the horizontal and tilted pool examination. As illustrated in Figure \ref{fig:pressure_solve}, our pressure solve with first order accuracy could not project velocity to a divergence-free field, however, we can acquire divergence-free velocity \fix{using} our second order pressure solve.
\subsection{Simulation results using moving grids}
\begin{figure}[htb]
    \centering
    \includegraphics[width=0.95\linewidth]{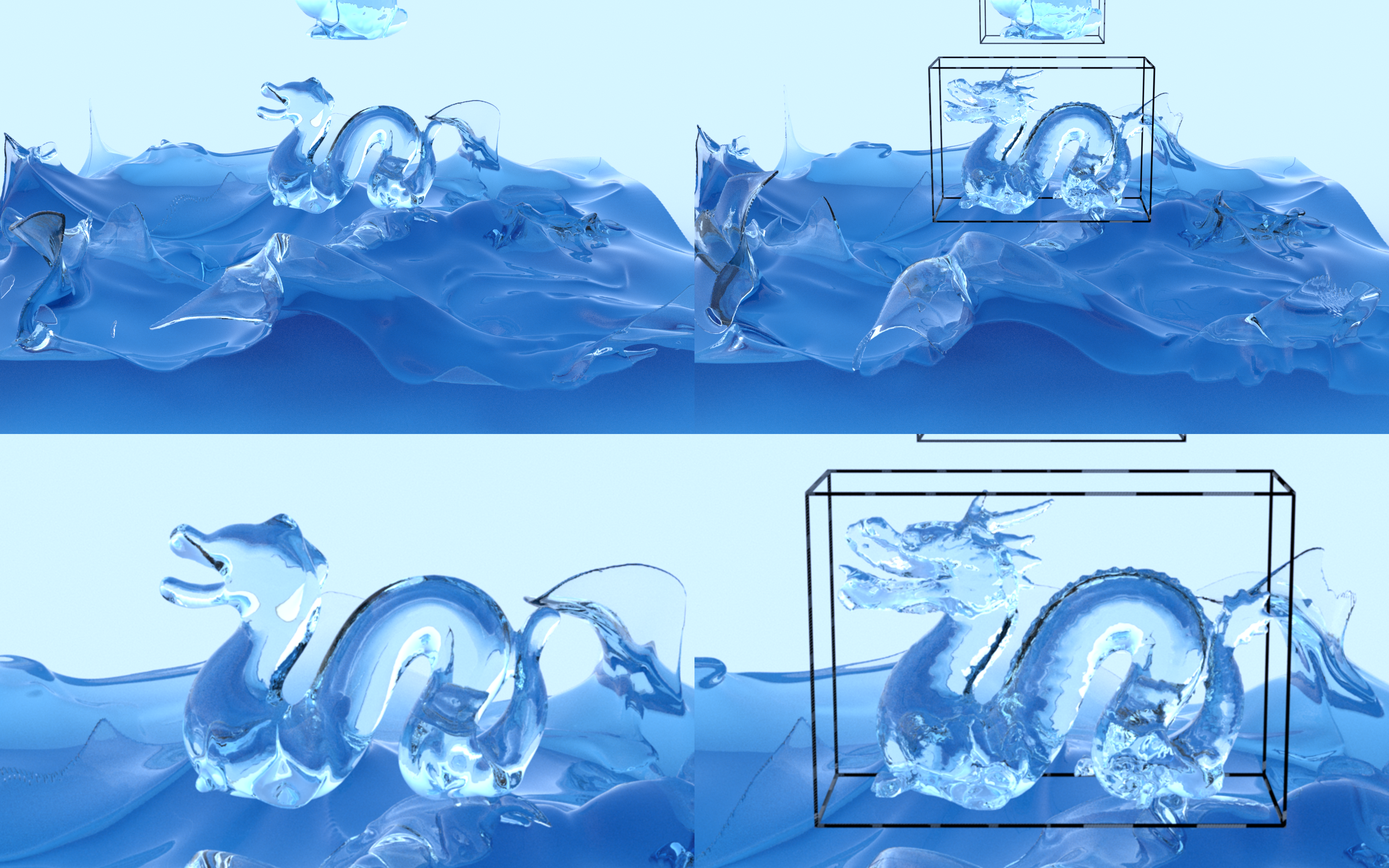}
    \caption[Comparison of the objects falling experiment]{\label{fig:dragon_fall}
        Comparison of previous and proposed methods on liquid objects \ry{falling into a pool simulated} with \ry{grids of} $600^3$ \ry{resolutions}. Left: previous method. \ry{Right:} our method.}
\end{figure}

% ours300 6.20, 37.21, 1.69, 46.07
% prev300 4.38, 35.48, n/a, 40.67
\begin{table}[htb]
    \centering
    \begin{tabular}{|c|c|c|c|c|} \hline
         & \multicolumn{2}{c|}{Fig.\ref{fig:dragon_fall}} & \multicolumn{2}{c|}{Half Reso of Fig.\ref{fig:dragon_fall}} \\ \cline{2-5}
        & Previous & Ours & Previous & Ours \\ \hline
        Advection & 92.01 & 102.05 & 4.27 & 6.20 \\ \hline
        Projection & 1080.55 & 1097.69 & 35.48 & 37.21  \\ \hline
        Update moving grids & N/A & 32.32 & N/A & 1.69 \\ \hline
        Total & 1188.76 & 1249.69 & 40.67 & 46.07 \\ \hline
    \end{tabular}
    \caption{Timings (in second) per frame of previous and our method in the scene of Figure \ref{fig:dragon_fall} and that with half resolution.}
    \label{tab:timing_obj_rain}
\end{table}
We ran some experiments using our method and compared them with the previous method \cite{FF01}. Table \ref{tab:timing_obj_rain} shows the computational time in the scene of Figure \ref{fig:dragon_fall}.
%
%Figure \ref{fig:crown_splash} shows the comparison in the water drop test. It can be seen that the shape of the water crown splash is clearer with our method. In this experiment, we set moving grids as shown in Figure \ref{fig:water_drop_mg_setting}. As for the computational run time, it is 1.18 times longer, and in particular, the interpolation used in advection takes almost 3 times longer. We discuss about this point in Section \ref{chap:discussion}.
%
\par
Figure \ref{fig:dragon_fall} shows the results of the liquid objects falling experiment. In this test, we set moving grids with the same speed of the falling objects. This prevents missing details in the objects. From this result, we can see that our method with moving grids is able to suppress the effect of numerical diffusion.
The computational time of our method is $1.05$ times longer than the \fix{standard liquid solve without our technique}, and in particular, the advection part takes $1.11$ times longer. When it comes to the half resolution results, our method took $1.13$ times longer for the computational time, and $1.45$ times longer for the advection time. We discuss this point in Section \ref{chap:discussion}.
\begin{figure}[htb]
    \centering
    \subfigure{\includegraphics[width=0.55\linewidth]{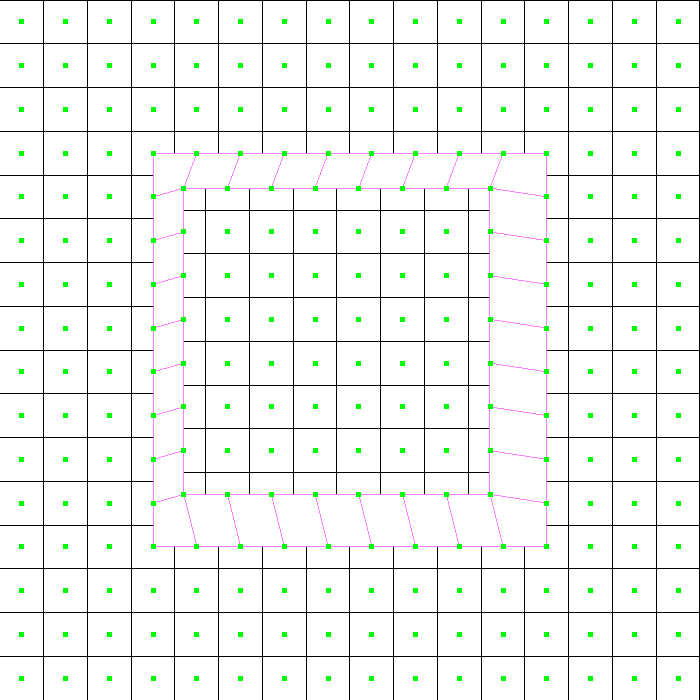}}
    \subfigure{\includegraphics[width=0.49\linewidth]{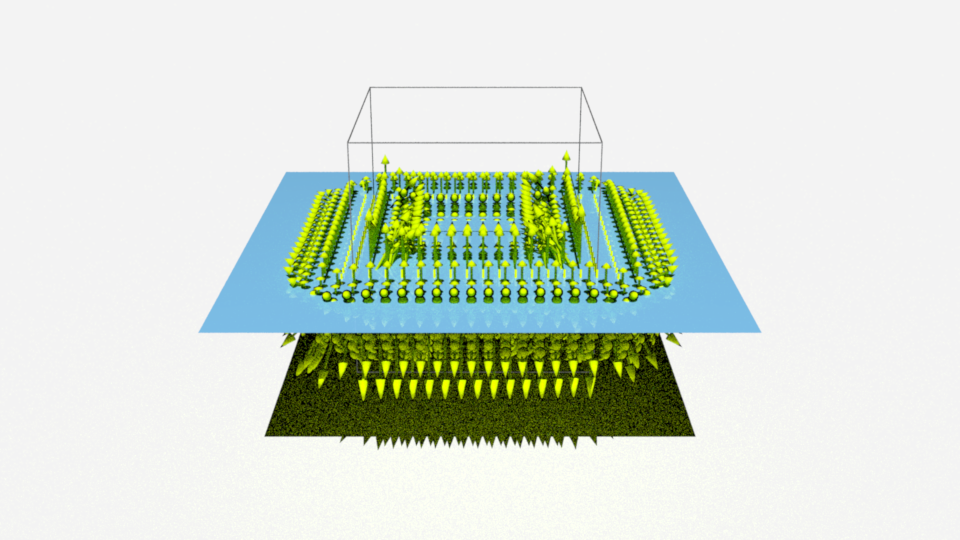}}
    \subfigure{\includegraphics[width=0.49\linewidth]{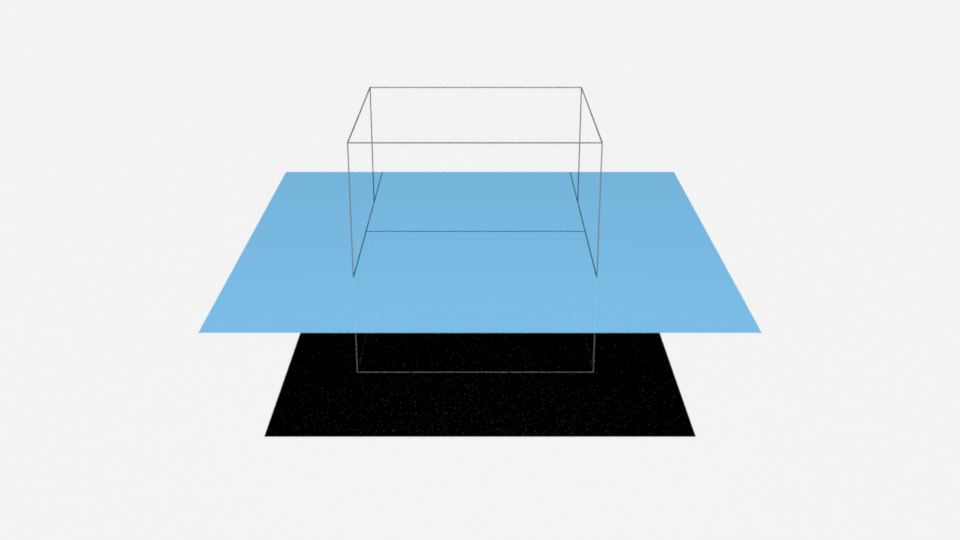}}
    \subfigure{\includegraphics[width=0.49\linewidth]{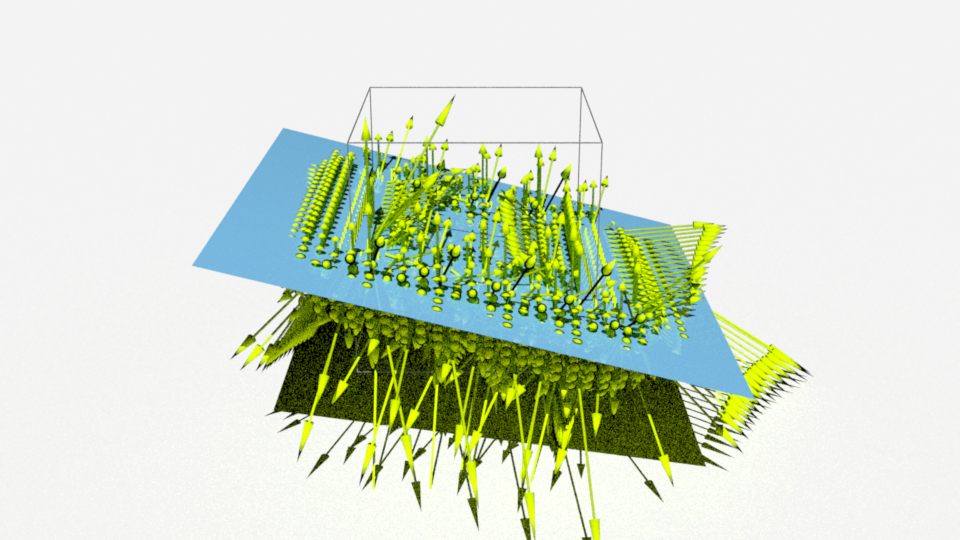}}
    \subfigure{\includegraphics[width=0.49\linewidth]{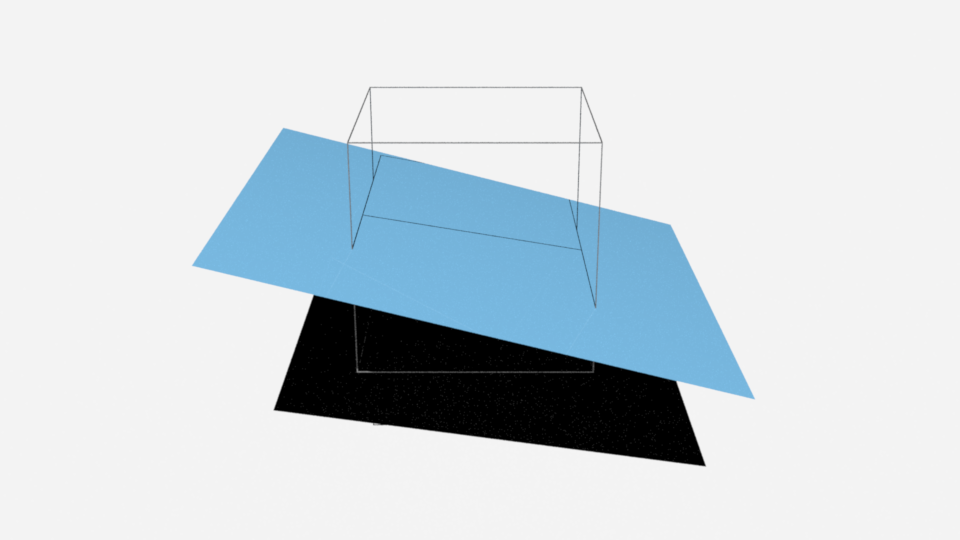}}
    \caption[Pressure solver test on the horizontal and tilted pool]{\label{fig:solver_test}
        Results of our pressure solver test on the horizontal pool and tilted pool. Top one illustrates the simple visualization of grid geometry in 2D. We add strong gravity force in the direction normal to the liquid surface. Yellow arrows represents velocity after projection. Left: naive first order projection. Right: second order projection(SPD).}
\end{figure}
\clearpage
\section{Discussion and Conclusion}\label{chap:discussion}

This paper presented a new Eulerian simulation framework for liquid simulation that leverages both finite element and finite volume methods, and we showed the numerical robustness of our MLS-based interpolation scheme and our new pressure solve. In addition, we \fix{showed} the applicability of our method with moving grids, and \fix{showed} that it is possible to suppress numerical diffusion.
\par
As shown in the interpolation test of the quadratic field in Figure \ref{fig:error_cell} and \ref{fig:error_face}, even with trilinear interpolation, the error becomes \fix{worse} \fix{as evaluation points get farther} from the defining \fix{points}.
\fix{This indicates that covering} the grid with \fix{moving grids} of the same velocity as the liquid achieves \fix{that} interpolation \fix{takes place} near the defining point, which \fix{results in} reduced error.
\par
\begin{figure}[htbp]
    \centering
    \subfigure{\includegraphics[width=0.49\linewidth]{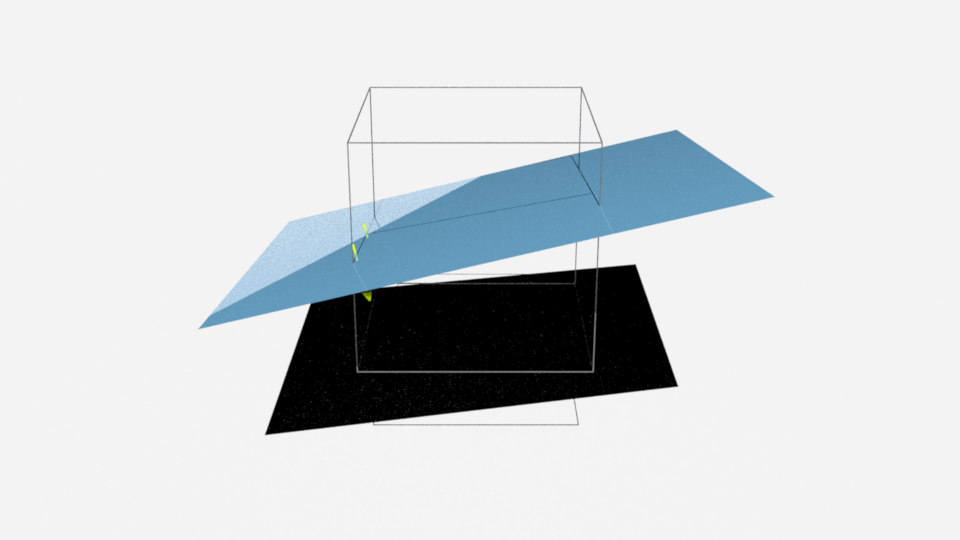}}
    \subfigure{\includegraphics[width=0.49\linewidth]{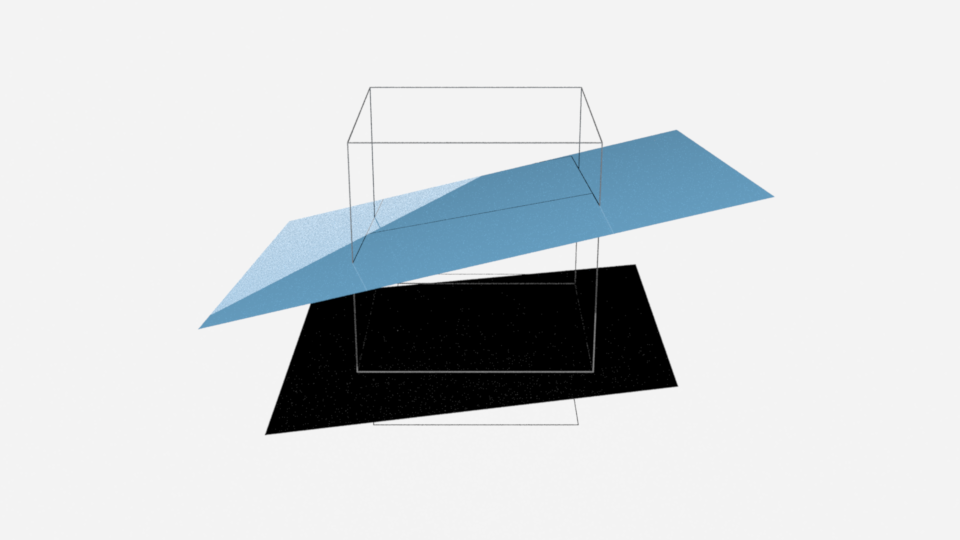}}
    \caption[Comparison of SPD and second order accuracy]{\label{fig:SPDvs2nd}
        Comparison of SPD and second order accuracy. In the tilted pool, we observed some errors with SPD, but not \fix{for} the case of imposing full second order accuracy.}
\end{figure}

As for the pressure solver, we found that our \fix{projected} SPD resulted in some errors, as shown in Figure \ref{fig:SPDvs2nd}. However, as explained in previous work \cite{ando2020}, using the SPD matrix is a satisfactory alternative because it is stable and \fix{fast to converge}.
\par
Although our method is currently effective in suppressing numerical diffusion, it may also be effective with respect to reducing computational run time if the relative velocity can be used for optimization when computing time step under the CFL condition.
\par
As for the performance, we are currently \fix{on the experimental stage, and as such,} we \fix{did not} optimize the code \fix{for faster performance}. In particular, for advection, the search for MLS interpolation sample points has not been optimized, which results in many unnecessary calculations. However, since our method restricts the use of FEM to moving grid boundaries, the computational overhead is relatively small as the FEM portion decreases with increasing resolution. Of course, optimization is important, but the low overhead of using FEM is one of the advantages of our method.
\par
In our experiment, we employed hexahedral elements, however, we believe that tetrahedral elements can be used as well \fix{at the cost of compromised} the accuracy of the linear interpolation.
\par
In the future work, it will be more meaningful if the moving grids region can be adaptively determined and if rotational motion can be considered. We are also considering the possibility of using finite elements for \fix{dealing with the complexity arising from} T-junction \fix{in} octree grids \cite{ando2020}.

%-------------------------------------------------------------------------

\section*{Acknowledgments}
This research was supported by the JSPS Grant-in-Aid for Young Scientists (18K18060).

%Bibliography
\bibliographystyle{unsrt}  
\bibliography{references}  

\end{document}